\documentclass[twocolumn,aps,prb,superscriptaddress,longbibliography]{revtex4-2}
\usepackage{graphicx}
\usepackage{color}
\usepackage{amsmath}
\usepackage{enumitem}
\usepackage{amssymb}
\usepackage{hyperref}
\usepackage[normalem]{ulem}
\usepackage{adjustbox}
\usepackage{comment}
\hypersetup{
	colorlinks = true,
	linkcolor = blue,
	citecolor = blue,
	urlcolor  = blue,
}
\usepackage[inkscapeformat=png]{svg}
\usepackage{caption}
\usepackage{subcaption}

\newcommand{\be}{\begin{equation}}
	\newcommand{\ee}{\end{equation}}

\newcommand{\bea}{\begin{eqnarray}}
	\newcommand{\eea}{\end{eqnarray}}

\renewcommand{\vec}[1]{{\boldsymbol #1}}
\renewcommand{\epsilon}{\varepsilon}

\begin{document}
	\title{Collapse of Landau level in semi-Dirac materials}
	\date{\today}
	
	\author{Daniil Asafov}
	\affiliation{California Institute of Technology, Pasadena, California 91125, USA}
	
	\author{Ilia Pavlov}
	\affiliation{School of Physics and Engineering,
ITMO University, 197101 St. Petersburg, Russia}
	
	\begin{abstract}
		Two-dimensional semi-Dirac models describe a family of novel materials that have anisotropic dispersion with relativistic-type spectrum along one of the spatial directions and non-relativistic along another. In the present paper we perform a detailed analysis of Landau levels collapse for such models in perpendicular magnetic field and in-plane electric field by using semi-classical approach and tight-binding simulations. The anisotropic nature of semi-Dirac dispersion manifests itself in the possibility of Landau level collapse for only one of the electric field directions. In addition, the topological transition with merging Dirac cones has its own distinct features in Landau level collapse.  
	\end{abstract}
	\maketitle
	
	\section{Introduction}

The topological transition in novel two-dimensional materials with two Dirac cones moving and merging at one point in momentum space attracted great attention in the literature \cite{Hasegawa2006PRB,Montambaux2009PRB,Montambaux2009EPJ}. The corresponding systems are called semi-Dirac systems and support quasiparticles with effective Hamiltonian having relativistic-type linear dispersion along one direction in momentum space and non-relativistic quadratic dispersion along another direction. 
Several effective semi-Dirac models were used to describe physics in the black phosphorus \cite{Ezawa2015,Pyatkovskiy2016PRB,Carbotte2019,Carbotte2019PRB,Adroguer2016,Jang2019}. In the simplest versions of semi-Dirac models the additional gap parameter controls whether two bands will be separated, touching in one point or two Dirac cones will appear \cite{Illes2015,Illes2016,Carbotte2019,Carbotte2019PRB,Oriekhov2022semiDirac}.There is a number of experimental realizations of such systems in 
optical lattices \cite{Tarruell2012Nature} and in microwave cavities \cite{Bellec2013PRL}. Typically the parameters of setup can be tuned to demonstrate different types of quasiparticle spectrum in the same system. Many physical properties of semi-Dirac materials were studied in recent years: among them the optical conductivity for different types of semi-Dirac systems \cite{Adroguer2016,Ziegler2017,Mawrie2019,Carbotte2019,Carbotte2019PRB,Jang2019,Carbotte2019,Carbotte2019PRB},  the magneto-conductivity \cite{Zhou2021PRB, Sinhaprbmagnetoconductivity}, thermoelectric response \cite{Mandalarxiv} Hall conductivity \cite{Sinhaprbnanoribbon} and the unusual scaling of Landau levels with magnetic field \cite{Montambaux2009EPJ,Montambaux2009PRB}. 

In the present paper we concentrate our attention on phenomena of collapse of Landau levels in an in-plane electric field for the semi-Dirac system. Firstly the presence of such phenomenon was discovered for relativistic quasiparticles in graphene \cite{Lukoze2007,Peres2007}. It was found that When the electric field strength approaches certain critical value, the formal solutions for Landau levels becomes singular. The gap-separated Landau levels are expected to disappear, and the continuos spectrum is instead formed. Such effect can be also understood in terms of Lotentz transformation, when the effective moving frame speed coincides with the speed of massless quasiparticles \cite{Lukoze2007,SHYTOV20091087,Nimyi2022PRB}.
Later the Landau level collapse was observed experimentally by measuring the field when Shubnikov-de-Haas oscillations disappear for the given sample \cite{Gu2011PRL}.
Later it was found, that also radial electric field can lead to similar collapse effect \cite{Nimyi2022PRB}.

	The is a number of differences in collapse of Landau levels that are expected for semi-Dirac system comparing to more typical Dirac semimetals. The main difference is caused by anisotropic structure of the spectrum, which itself should lead to anisotropy in critical values of electric field. In addition, a set of two quasiclassically disconnected parts of Fermi surface is present in the cases of two separated Dirac cones \cite{Montambaux2009EPJ,Montambaux2009PRB,Carbotte2019PRB}, and one may expect an additional effect of tunneling that will modify Landau level collapse in such a case. The open question of how Landau level collapse happens in semi-Dirac models motivate the study in the present paper.
	
	The paper is organized as follows:
	in Sec.\ref{sec:semi-Dirac-recall} we recall the main properties of low-energy semi-Dirac model with gap parameter that controls topological phase transition. Next, in Sec.\ref{sec:semi-Dirac-semi-classics} we present a semi-classical calculations of Landau levels that appear in crossed electric and magnetic field. Additionally, we present asymptotic analysis for exact Schr\"{o}dinger equation for such system. The special attention is given to the case when the tunneling (also called magnetic breakdown) takes part and modifies the structure of quasiclassical Landau levels. The corresponding calculations are presented in Sec.\ref{sec:semi-Dirac-semi-classics-MB}. Finally, in Sec.\ref{sec:conclusions} we compare the results with those known for graphene and give concluding remarks.

	\section{Semi-Dirac model and topological transition with merging Dirac cones}
	\label{sec:semi-Dirac-recall}
	The effective Hamiltonian that captures separated Dirac cones and gapped regime as well as topological transition between them is given by the following expression:
	\begin{align}
		\label{eq:universal-Hamiltonian}
		H=(\Delta+a k_x^2)\sigma_x+v k_y \sigma_y.
	\end{align}
	There are a number of modification of this Hamiltonian with higher-order terms or additional gap parameters with another matrix structure \cite{Jang2019}. Here we analyze the simplest version of such Hamiltonian, which was called `universal' in Refs.\cite{Montambaux2009EPJ,Montambaux2009PRB} as it captures the essential physics of Landau level collapse. The dispersion for each of two bands in the absence of external fields in given by
	\begin{align}
		\epsilon_{\pm}(\vec{k})=\pm \sqrt{(a k_x^2+\Delta)^2+v^2 k_y^2}.
        \label{eq:semi-Dirac-effective-dispersion}
	\end{align}
    The two band-touching Dirac points are separated by the distance $\delta_x=2\sqrt{\Delta/a}$ in momentum space along x-direction in the case of negative $\Delta<0$ assuming $a,v>0$. In the case of zero gap parameter $\Delta=0$ the two bands touch at $\vec{k}=0$. 
	
	\section{Semi-classical quantization of cyclotron orbits and magnetic breakdown}
	\label{sec:semi-Dirac-semi-classics}
\begin{figure}
     \centering
     \includegraphics[width=\linewidth]{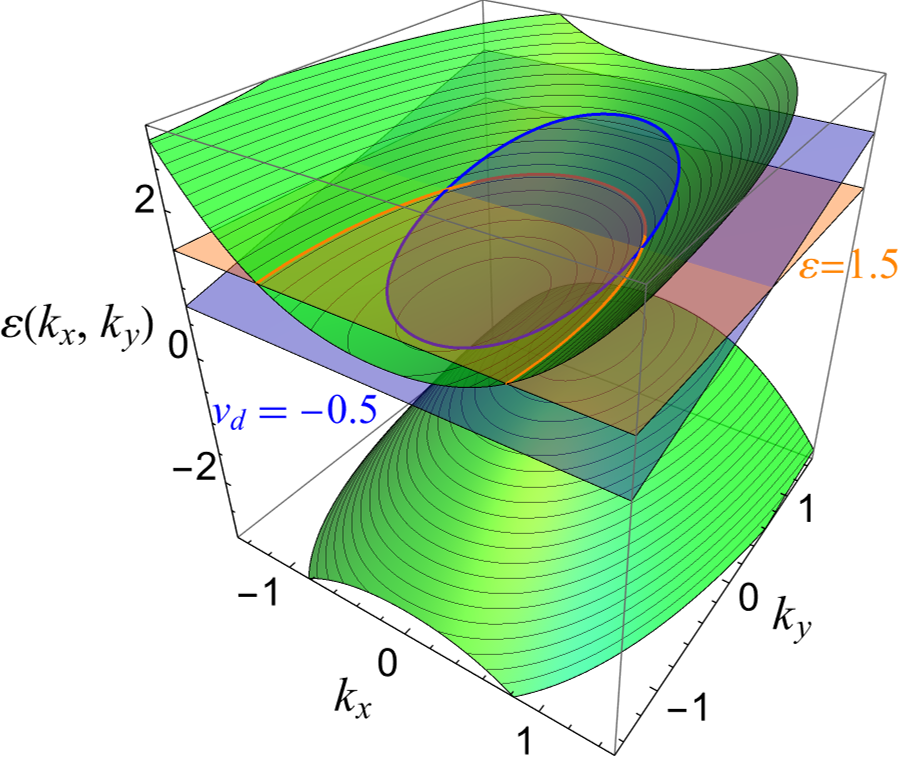}
     \caption{Spectrum given by the equation \ref{eq:semi-Dirac-effective-dispersion}. The gap parameter is $\Delta=1$. We choose units $v=1,\, a=1$. Two planes on the picture represent additional $\vec{v}_d \vec{k}$ part from the \ref{Peierls-Osnager}. As illustrated by the blue plane, when $v_d$ is the small intersection of the plane and the spectrum figure is a closed orbit and thus, Landau-level exists. However, when $v_d$ is very large, as with the orange plane, the intersection is an open orbit and therefore we can say that we observed the collapse of Landau levels.}
     \label{fig:geom_interp}
 \end{figure}

	The description of Landau levels in the effective model given by Hamiltonian \eqref{eq:universal-Hamiltonian} was presented in Ref.\cite{Montambaux2009EPJ}. The main features such as scaling with magnetic field and Landau level index can be captured by using Peierls-Onsager quasiclassical quantization rule, which states that the area of the contant energy curve in momentum space should be quantized 
	\begin{align}
		S(\epsilon(\vec{k})=const)=2\pi(N+\gamma)e B.
	\end{align}
	 As was pointed out in \cite{IMLifshitz_1960}, the presence of electric field modifies the Peierls-Onsager quantization rule:
	 \begin{align}
	 	S(\epsilon(\vec{k})-\vec{v}_d \vec{k}=const)=2\pi(N+\gamma)e B,
	 \label{Peierls-Osnager}
	 \end{align}
	with classical drift velocity in crossed fields defined as $\vec{v}_{d}=c \vec{E}\times\vec{B} / \vec{B}^2$.  Before proceeding with more detailed calculations let us perform qualitative analysis. The collapse of Landau levels is associated with the fact that the area defined by contour in momentum space $\epsilon-\vec{v}_d \vec{k}=const$ becomes infinite. This may happen either for Landau level with particular index or for all levels simultaneously. Consequently, the general criteria for the presence of Landau level collapse for the band spectrum in infinite momentum space $-\infty<k_{x,y}<\infty$ is equivalent to the question of existence of closed curve of solutions for equation $\epsilon(\vec{k})-\vec{v}_d \vec{k}=const$ with fixed right-hand side constant. The geometrical interpretation of this criteria can be formulated in terms of  shape of cross-section of the dispersion $\epsilon(\vec{k})$ by the plane $\vec{v}_d \vec{k}+const=0$, as illustrated on the figure  [\ref{fig:geom_interp}].
Such cross-section becomes an open curve when the dispersion grows slower than linear with pre-factor $\vec{v}_{d}$ at large momenta values:
	\begin{align}\label{eq:simple-criteria}
		\epsilon(\vec{k}\to\infty)\lesssim \vec{v}_d \vec{k}.
	\end{align}
    This formula allows one to predict the existence of Landau level collapse by analyzing the asymptotic behavior of dispersion. 
     
	Applying \eqref{eq:simple-criteria} to the dispersion of semi-Dirac model and taking asymptotics along two orthogonal directions,
	\begin{align}
		\epsilon_{+}(k_x\to\infty)\sim  a k_x^2,\quad \epsilon_{\pm}(k_y\to\infty)\sim \pm v k_y,
	\end{align}
we find that collapse of Landau level is possible for the case of drift velocity having a nonzero component along the y-direction. The collapse happens for all energy levels simultaneously when this component is bigger than the parameter of the model $v_{d,y}>v$. For the case of the uniform constant electric and magnetic field with B-field along the z-direction, this leads to the fact that the electric field should have a nonzero component along $x$ direction, with critical value for the collapse:
\begin{align}
	E_{x,c}=v B_z / c.
\end{align}
For the electric field directed along $y$-axis of the model, the collapse of Landau levels is not possible. 
\newline
\section{Energy levels in the presence of electromagnetic field}
\begin{figure}
    \includegraphics[width=\linewidth]{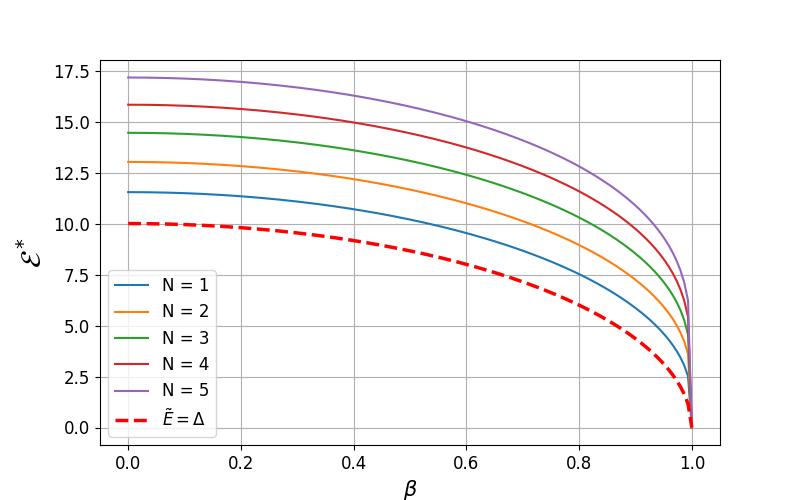}
    \caption{Numerical solution for the equation \ref{eq:energy levels D+} with $\Delta=mv^2_F=\frac{eB}{mc}$. Here $\beta=\frac{v_d}{v_F}$ and $\mathcal{E^*}$ is measured in the units of $\frac{eB}{mc}$ }
    \label{fig:delta>0}
\end{figure}

\begin{figure}
     \centering
     \includegraphics[width=\linewidth]{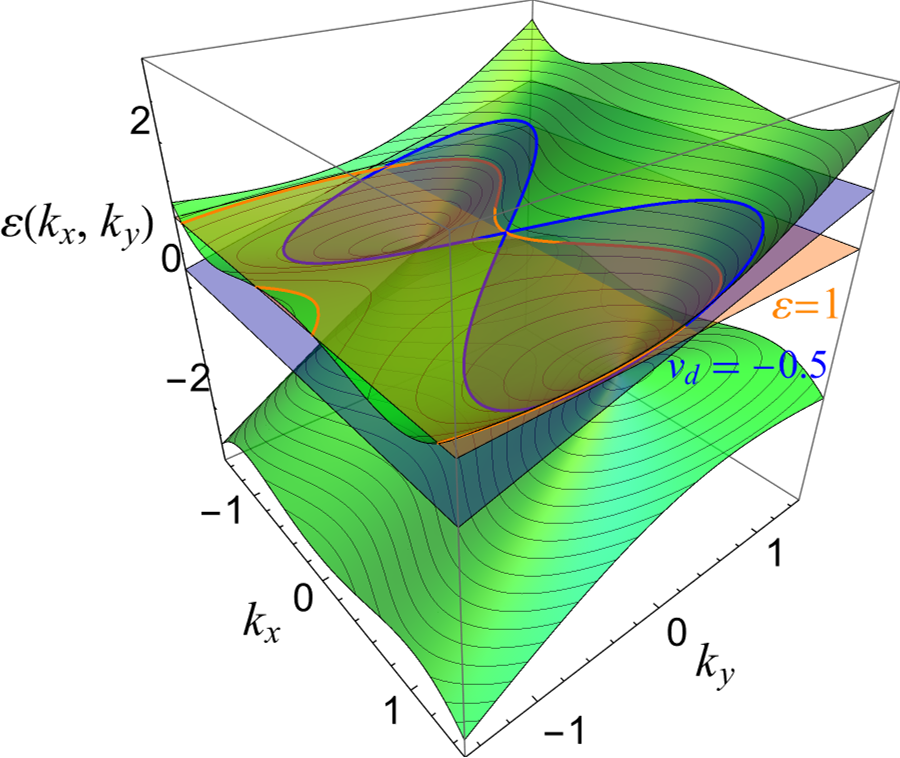}
     \caption{Spectrum given by the equation \ref{eq:semi-Dirac-effective-dispersion} with the parameter $\Delta=-1$. We notice that by regulating the electrical field strength two independent closed figures can emerge.}
     \label{fig: D<0}
\end{figure}

To calculate energy levels we will use the equation \ref{Peierls-Osnager}. Full technical calculations are given in the appendix \ref{appendix:integrals}, but the main idea is to notice, that the resulting closed orbit could be described by the equation
\begin{equation}
    (\frac{k^2_x}{2m} + \Delta)^2 + (v^2_F-v^2_d) (k_y - \frac{v_d \mathcal{E^*}(\vec{p})}{v^2_F-v^2_d})^2 = \frac{v^2_F}{v^2_F-v^2_d} \mathcal{E^*}^2
    \label{eq:closed orbit}
\end{equation}
Denoting $\Tilde{E} =\frac{v_F}{\sqrt{v_f^2-v_d^2}}\mathcal{E^*}$ and $(k_y - \frac{v_d \mathcal{E^*}(\vec{p})}{v^2_F-v^2_d})$
we can express the area of the orbit as 
\begin{equation}
    S(k_x, k_y) = \int \int dk_x d\Tilde{k}_y \theta\left(\Tilde{E}^2 -  (\frac{k^2_x}{2m} + \Delta)^2 - (v^2_F-v^2_d)\Tilde{k}^2_y\right)
    \label{eq:surface equation}
\end{equation}
In the case of $\Delta \geq 0$ the equation for energy levels has the form
\begin{equation}
   \frac{\pi(|\Tilde{E}|-\Delta)\sqrt{|\Tilde{E}|+\Delta}}{4\sqrt{a}\sqrt{v^2-v_d^2}}F(\frac{1}{2}, \frac{3}{2}, 3, - \frac{|\Tilde{E}|-\Delta}{|\Tilde{E}|+\Delta}) = 2\pi (N+\gamma) \frac{eB}{c}
   \label{eq:energy levels D+}
\end{equation}.
This equation can be solved numerically. On the figure \ref{fig:delta>0} we plot the solution for a particular $\Delta$, but the whole form is quite general for the case $\Delta \geq 0$.
\bigskip
\newline
In the case of negative gap parameter $\Delta<0$ the effective dispersion \eqref{eq:semi-Dirac-effective-dispersion} contains two Dirac points separated in momentum space. As a result, there are two closed "orbits" possible (see figure \ref{fig: D<0})

\begin{figure}
     \centering
     \includegraphics[width=0.8\linewidth]{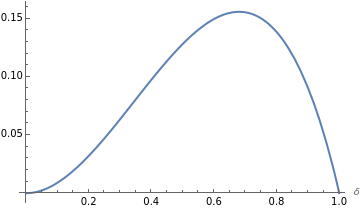}
     \caption{Plot of the function $f(\delta)$}
     \label{fig: function}
\end{figure}

In this situation there is a possibility of tunneling between two orbits which influences the energy levels structure. The tunneling probability decays exponentially with the energy deviation from the saddle point level. Thus, we should implement double-well representation to properly describe the semi-classical spectrum around that point. The complete theory of corresponding semiclassical description was presented in Ref.\cite{Alexandradinata2018}. Here we use the quantization condition derived from this description:
	\begin{align}
		\cos\left[\frac{\Omega_1+\Omega_2}{2}+\phi(\mu)\right]=|\mathcal{T}(\mu)|\cos\left[\frac{\Omega_1-\Omega_2}{2}\right].
	\end{align} 
 $\Omega_j= \pi + \frac{eB}{c}S_j + \int_0^1 \frac{\Tilde{H}^{\nu(t_j)}_1}{v^x_{\nu(t_j)}}\frac{dk}{dt_j} \,dt_j + (-1)^{j+1}\delta^B_y$.
 $S_j$ - is the area of the corresponding orbit, $\Tilde{H_1}=H^R_1+H^Z_1+H^B_1 - H_1(0)$ - correspond to Roth, Zeeman and Berry corrections and $\delta^B_y$ is an additional Berry phase. Since in our case Berry curvature is equal to zero and the Zeeman Hamiltonian is absent the only term left is the area of the orbit in $k$-space $S_j$. Noticing that the areas of the orbits are identical, we get the expression:
 \begin{align}
		\cos\left[\frac{c}{eB}S+\phi(\mu)\right]=-\frac{e^{\pi \mu/2}}{\sqrt{2\cosh{\pi \mu}}}.
  \label{quantonisation with negative delta.1}
\end{align}
The area of an orbit can be calculated by the same method we used earlier. Denoting $\delta = |\frac{\tilde{E}}{\Delta}|$ and using the equation  \ref{eq:surface equation} we obtain
\begin{equation}
    S = \frac{\sqrt{2m} |\Delta|^{\frac{3}{2}} \delta^2 \sqrt{1-\delta}}{2\sqrt{v^2_F-v^2_d}} F(\frac{1}{2}, \frac{3}{2}, 3, -\frac{2 \delta}{1-\delta})
    \label{eq: area for minus delta}
\end{equation}
Using the definition of $\phi$ and $\mu$ from the \cite{Alexandradinata2018}
\begin{equation}
    \phi (\mu) = arg[\Gamma(\frac{1}{2}-i\mu)] + \mu( \ln(|\mu|) - 1)
\end{equation}
\begin{equation}
 \mu = \pm \sqrt{\frac{2m}{v^2_F-v^2_d}} \frac{c|\Delta|^{\frac{3}{2}}}{eB}(1-\delta)
\end{equation}
Where $\mp$ corresponds to the branch choice in the equation \ref{eq:semi-Dirac-effective-dispersion}. Solving this final equation numerically, we get the structure of energy levels described on the figure \ref{fig: delta<0}.
\begin{figure}
\centering
\begin{subfigure}{0.45\textwidth}
    \includegraphics[width=\textwidth]{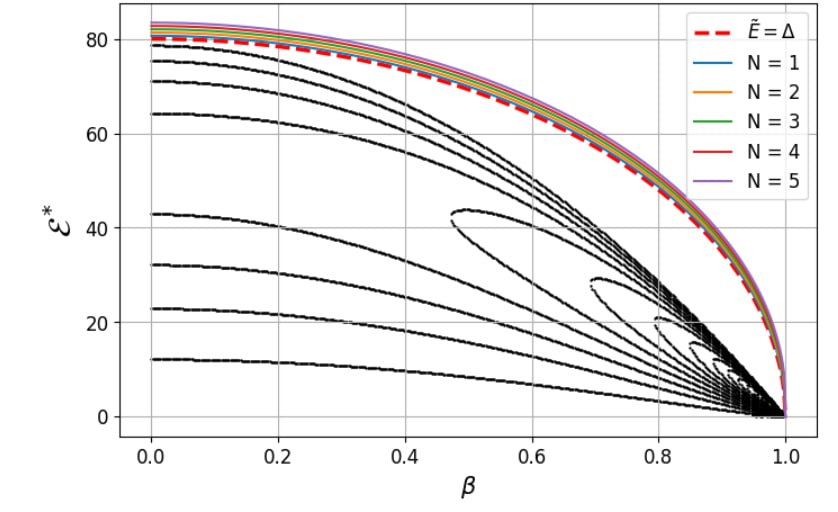}
    \caption{$\Delta=-\frac{1}{4}\frac{eB}{mc}$}
    \label{fig:first}
\end{subfigure}
\hfill
\begin{subfigure}{0.46\textwidth}
    \includegraphics[width=\textwidth]{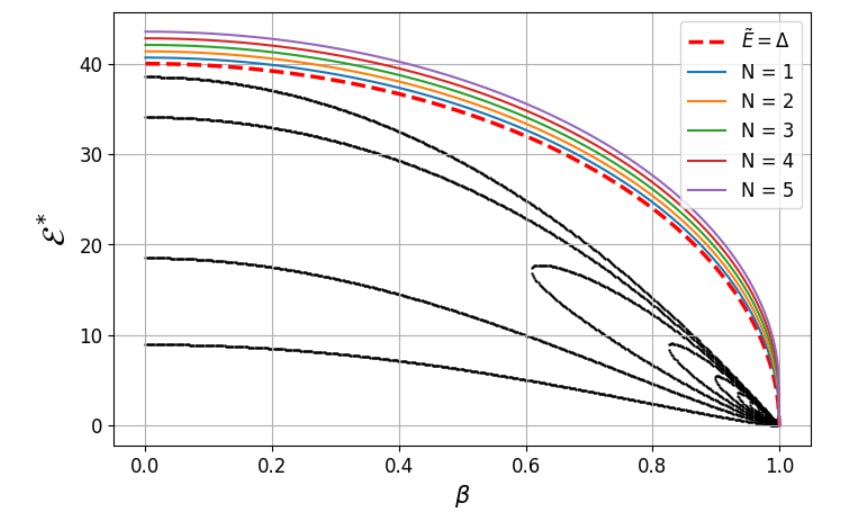}
    \caption{$\Delta=-\frac{1}{2}\frac{eB}{mc}$}
    \label{fig:second}
\end{subfigure}
\hfill
\begin{subfigure}{0.49\textwidth}
    \includegraphics[width=\textwidth]{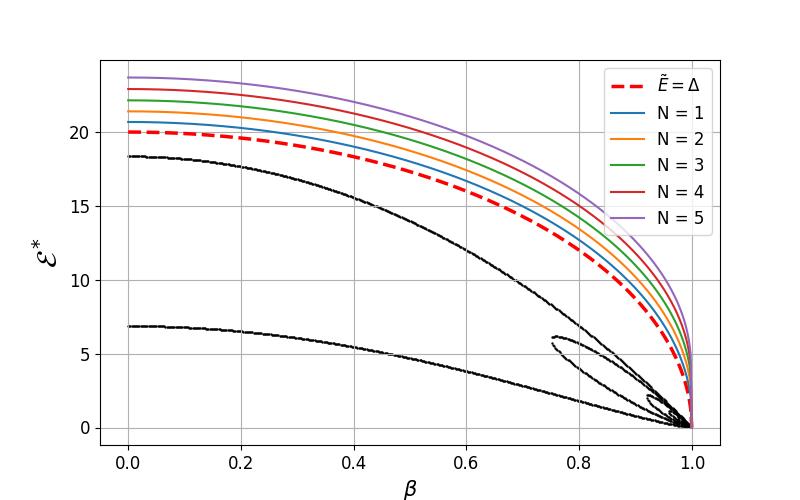}
    \caption{$\Delta=-\frac{eB}{mc}$}
    \label{fig:third}
\end{subfigure}
        
\caption{Numerical solution for the equation \ref{quantonisation with negative delta.1} $+$ in the equation \ref{eq:semi-Dirac-effective-dispersion}. High energy levels have the same structure as on the figure \ref{fig:delta>0}, however, below the level $\tilde{E}=|\Delta|$ we get a different structure, including the emergence of "loops". As before, $\beta=\frac{v_d}{v_F}$, $\Delta=-2mv^2_F$ and $\mathcal{E^*}$ is measured in the units of $\frac{eB}{mc}$. The magnetic field being increased two times from picture to picture. The stronger magnetic field is the less levels below the saddle point we have}
\label{fig:figures delta<0}
\end{figure}
One can notice, that the conditions for Landau-levels collapse does not depend on $\Delta$ and happens when $\beta=1$.
\bigskip
\newline
One can observe a peculiar effect: the birth of pairs of energy levels with the increase of the electrical field.
The mathematical origin of such a solution can be illustrated pretty straightforwardly.  
\newline
Let us take the limit $\mu \gg 1$ in the equation \ref{quantonisation with negative delta.1}, which arises at relatively small energies and large fields. In this limit, we can use the form of Stirling's approximation for complex arguments with large absolute value (see chapter IV in \cite{Zetafunc}) and expand $arg[\Gamma(\frac{1}{2}-i\mu)] \approx -\mu(\ln{\mu}-1)$. Then, the equation \ref{quantonisation with negative delta.1} turns into
\begin{equation}
    \cos\left[\frac{c\sqrt{2m} |\Delta|^{\frac{3}{2}}}{2eB\sqrt{v^2_F-v^2_d}} \delta^2 \sqrt{1-\delta} F(\frac{1}{2}, \frac{3}{2}, 3, -\frac{2 \delta}{1-\delta})\right]=-1
    \label{eq:large mu}
\end{equation}
Denoting $\frac{c\sqrt{2m} |\Delta|^{\frac{3}{2}}}{2eB\sqrt{v^2_F-v^2_d}}$ as $\tilde{\omega}$ we can rewrite the equation as
\begin{equation}
    \cos\left[\tilde{\omega} \delta^2 \sqrt{1-\delta} F(\frac{1}{2}, \frac{3}{2}, 3, -\frac{2 \delta}{1-\delta})\right] = -1
\end{equation}
Now, let's denote the function $f(\delta) = \delta^2 \sqrt{1-\delta} F(\frac{1}{2}, \frac{3}{2}, 3, -\frac{2 \delta}{1-\delta})$ and look at this function at the point $\delta=\delta_0$, such that $\frac{d}{d \delta} f(\delta) = 0 \Big{|}_{\delta=\delta_0}$. Figure \ref{fig: function} suggests that there is only one point like that. This point is also an extremum for the function $\cos\left[\tilde{\omega} f(\delta) \right]$. Moreover, this is an extremum of that function for any $\tilde{\omega}$. We can see that because the function $f(\delta)$ is continuously growing on the interval $(0, \delta_0)$ and is continuously decaying on the $(\delta_0, 1)$ different $\tilde{\omega}$ just correspond to different oscillation frequencies and thus two different number of times the cosinus function touches the line $f(\delta)=-1$. The new pair of levels is born whenever $\cos{(\tilde{\omega} f(\delta_0))}$ = -1. From here we can write the condition on birth of a pair of levels as
\begin{equation}
    \sqrt{1-\beta_*^2}=\frac{c\sqrt{2m} |\Delta|^{\frac{3}{2}}}{2eBv_F}\frac{f(\delta_0)}{\pi+2\pi n}
\end{equation}
Where $\beta_*$ are the values of $\frac{v_d}{v_f}$ at which new pairs of levels are born. Numerical calculations of $f(\delta_0)=0.155829$.
\newline
\bigskip
The observed appearance of pairs of energy levels with growing electric field can be understood in the following way: this phenomenon happens on the level of saddle point, where the tunneling effects are strong. For a typical Landau level the orbit remains closed with changing electric field, thus Landau level energy moves to the lower energy. In the case of Landau levels near saddle point the orbit might pass the saddle point level and break down. Such modified quantization rule might allow for the appearance of new allowed Landau levels when the effective dispersion $\mathcal{E}^*$ deforms under the growing electric field. 

\section{Conclusions}
To summarise, we have analyzed the semi-Dirac model subjected to in-plane electric and out-of-plane magnetic field and found the condition for the appearance of Landau level collapse phenomenon. Based on a semiclassical WKB-type approach, we found that the Landau level collapse appears only when the effective dispersion grows as linear function of momentum or slower. Application of this general criteria as well as exact derivation of semiclassical Landau level dispersion from the constant energy curves modified by electric field have shown that the LL collapse happens only for one direction of electric field. In other words, only electric field with the x-component exceeding the threshold leads to LL collapse. This result can be contrasted with the example of monolayer graphene, where any direction of electric field has possibility to make a collapse.

In addition, we analyzed the structure of LL for electric field values close to collapse for the gap parameter of the model corresponding to three topologically distinct cases (two Dirac cones, one band touching point and two gapped bands). We have discovered that in the case of the negative gap parameter $\Delta < 0$ the number of Landau levels below the saddle point increases.
 \section{Acknowledgments}
 We would like to express our gratitude to Leonid Levitov and Dmytro Oriekhov for the idea of the project and for their guidance and help. We would also like to thank Yelizaveta Kulynych for helpful discussions.
\label{sec:conclusions}	
\appendix
\onecolumngrid 
\section{Evaluation of area integrals for semi-classical quantization}
\label{appendix:integrals}

The effective dispersion of semi-Dirac model has the form
\begin{align}
\mathcal{E}(\vec{p})=\pm \sqrt{\left(\frac{k_x^2}{2m}+\Delta\right)^2+v^2_Fk_y^2}.
\end{align}
It predicts the relativistic dispersion along one direction and non-relativistic along another (and mix in between). The "unusual" power law comes from the anisotropy of the model and the corresponding constant energy curve area function with energy.
The Landau level collapse happens when electrical field has the component $E_x > \frac{v_f}{c}H$. The $E_y$ component has no effects on the collapse.
\newline
\bigskip
Let us find semiclassical energy levels in the presence of electrical field. The modified dispersion \cite{IMLifshitz_1960}
\begin{equation}
    \mathcal{E^*}(\vec{p})= \pm \sqrt{\left(\frac{k_x^2}{2m}+\Delta\right)^2+v^2_Fk_y^2} - v_d k_y = \textbf{const}
    \label{eq: effective dispersion with field2}
\end{equation}
In order to find areas in the momentum space it is convenient to rewrite this equation as
\begin{equation}
    (\frac{k^2_x}{2m} + \Delta)^2 + (v^2_F-v^2_d) (k_y - \frac{v_d \mathcal{E^*}(\vec{p})}{v^2_F-v^2_d})^2 = \frac{v^2_F}{v^2_F-v^2_d} \mathcal{E^*}^2
\end{equation}
From this equation it is clear the electrical field corresponds only to scaling of energy levels, and does not lead to other significant effects. It is worth noting that here we can see the collapse arising at $v_d = v_F$.
\newline
Let us denote in the following calculations that $\frac{v^2_F}{v^2_F-v^2_d} \mathcal{E^*}^2 = \Tilde{E}^2$ and $(k_y - \frac{v_d \mathcal{E^*}(\vec{p})}{v^2_F-v^2_d}) = \Tilde{k}_y$. The area of orbit in the momentum space can be calculated as 
\begin{equation}
    S(k_x, k_y) = \int \int dk_x d\Tilde{k}_y \theta(\Tilde{E}^2 -  (\frac{k^2_x}{2m} + \Delta)^2 - (v^2_F-v^2_d)\Tilde{k}^2_y)
    \label{eq:surface equation}
\end{equation}
There are two cases we need to study: $\Delta \geq 0$ and $\Delta <0$. Let us start with the first case. The integral comes down to 
\begin{equation}
     S(k_x, k_y) = \frac{2}{\sqrt{v^2_F-v^2_d}} \int_{-\sqrt{2m} \sqrt{|\Tilde{E}| - \Delta}}^{\sqrt{2m} \sqrt{|\Tilde{E}| - \Delta}} \sqrt{\Tilde{E}^2 -  (\frac{k^2_x}{2m} + \Delta)^2} \, dk_x\
\end{equation}
And can be transformed to
\begin{equation}
     S(k_x, k_y) = \frac{2 \sqrt{2m} (|\Tilde{E}|-\Delta)^{\frac{3}{2}}}{\sqrt{v^2_F-v^2_d}} \int_{-1}^{1} \sqrt{\frac{\Tilde{E}^2}{(|\Tilde{E}|-\Delta)^2} -  (x^2 + \frac{|\Tilde{E}|}{(|\Tilde{E}|-\Delta)} - 1)^2} \, dx\
\end{equation}
From here, changing the integration variable to $d (x^2)$ and using the integral representation of the hypergeometric function \begin{equation}
    F(a, b, c, z) = \frac{\Gamma(c)}{\Gamma(b) \Gamma(c-b)} \int_0^1 t^{b-1}(1-t)^{c-b-1}(1-zt)^{-a} \, dt\
    \label{eq:hypergeometric function}
\end{equation}
We get, that
\begin{equation}
 S(k_x, k_y) = \frac{2 \sqrt{2m} (|\Tilde{E}|-\Delta)\sqrt{|\Tilde{E}|+\Delta}}{\sqrt{v^2_F-v^2_d}} F(\frac{1}{2}, \frac{3}{2}, 3, -\frac{|\Tilde{E}| - \Delta}{|\Tilde{E}|+\Delta}) = 2\pi \frac{eB}{c}(n+\gamma)
 \label{eq: quantization with plus delta}
\end{equation}
\newline
\bigskip
When the $\Delta < 0$ an interband breakdown for some energy levels can arise, so our quantization rule should be modified. When 
$|{\frac{\Tilde{E}}{\Delta}}| = \delta \geq 1$ formula \ref{eq: quantization with plus delta} holds for negative $\Delta$ as well. However, when $\delta <1$ there exist two closed classical orbits and there is a possibility of tunnelling between them. This case is described in the paper \cite{Alexandradinata2018}
\begin{align}
		\cos\left[\frac{\Omega_1+\Omega_2}{2}+\phi(\mu)\right]=|\mathcal{T}(\mu)|\cos\left[\frac{\Omega_1-\Omega_2}{2}\right].
  \label{quantonisation with negative delta}
\end{align}
In our case, with the absence of Berry's curvative and Zeeman hamiltonian $\Omega_1$ and $\Omega_2$ are just equal to the dimensioneless surfaces of both trajectories in the momentum space. 
As we can see from the equation \ref{eq:surface equation} the Surfaces are equal to each other. Using the equation \ref{eq:surface equation} we obtain four possible "stop points" - the integration limits for surface calculation: $k_x=\pm \sqrt{2m(|\Delta| \pm |\Tilde{E}|)}$. Transforming to dimensioneless variables we get
\begin{equation}
     S_1=S_2 = \frac{2 \sqrt{2m} |\Delta|^{\frac{3}{2}}}{\sqrt{v^2_F-v^2_d}} \int_{-\sqrt{1-\delta}}^{-\sqrt{1+\delta}} \sqrt{\delta^2-(x^2-1)^2} \, dx\
\end{equation}
Changing the integration variable to $\frac{x^2-1}{\delta}$ and expanding the expression as $(\delta+1-x^2)(\delta+x^2-1)$ we will get
\begin{equation}
    S = \frac{2 \sqrt{2m} |\Delta|^{\frac{3}{2}} \delta^2}{\sqrt{v^2_F-v^2_d}} \int_{-1}^{1} \sqrt{(1-x)(1+x)}(1+\delta x) \, dx\
\end{equation}
From here it is easy to obtain the integral representation of a hypergeometric function \ref{eq:hypergeometric function} by changing the integration variable to $\frac{1+x}{2}$ and as a result we get
\begin{equation}
    S = \frac{\sqrt{2m} |\Delta|^{\frac{3}{2}} \delta^2 \sqrt{1-\delta}}{2\sqrt{v^2_F-v^2_d}} F(\frac{1}{2}, \frac{3}{2}, 3, -\frac{2 \delta}{1-\delta})
    \label{eq:area for minus delta1}
\end{equation}
$\mu$, $\phi(\mu)$ and $\mathcal{T}(\mu)$ we take as defined in the paper \cite{Alexandradinata2018}
\begin{equation}
 \mu = \pm \sqrt{\frac{2m}{v^2_F-v^2_d}} \frac{c|\Delta|^{\frac{3}{2}}}{eB}(1-\delta)
\end{equation}
Where choice of the sign corresponds to the brunch choice in the equation \ref{eq: effective dispersion with field2} (Note, that equation \ref{eq:area for minus delta1} includes both brunches due to the modulus of $\Tilde{E}$).
\begin{equation}
    \phi (\mu) = arg[\Gamma(\frac{1}{2}-i\mu)] + \mu( \ln(|\mu|) - 1)
\end{equation}
\begin{equation}
    \mathcal{T}(\mu) = e^{i\phi(\mu)}\frac{e^{\pi \mu/2}}{\sqrt{2\cosh{\pi \mu}}}
\end{equation}
And the only thing we left to do is to put them into this final equation
\begin{align}
		\cos\left[\frac{c}{eB}S+\phi(\mu)\right]=-\frac{e^{\pi \mu/2}}{\sqrt{2\cosh{\pi \mu}}}
  \label{quantonisation with negative delta}
\end{align}
\newline
\bigskip
An approximate solution of this equation could be found with the right estimation for $\text{arg}[\Gamma(\frac{1}{2}-i\mu)]$. 
To find this estimation, we use the Weierstrass's definition of the Gamma function
\begin{equation}
    \Gamma(z) = \frac{e^{-\gamma z}}{z}\prod_{n=1}^{\infty} (1+\frac{z}{n})^{-1}e^{z/n}
    \label{Weierstrass's definition}
\end{equation}
Where $\gamma$ is the Euler–Mascheroni constant
\newline
Let's define the phase $\phi_n$ as $(z+n)=|z+n|e^{i\phi_n}$.
Then, according to \ref{Weierstrass's definition}
\begin{equation}
\text{arg}[\Gamma(\frac{1
}{2}-i\mu)] = \gamma \mu - \phi_0 -\sum_{n=1}^{\infty} (\frac{\mu}{n}+\phi_n)
\end{equation}
Where we can take $\phi_n = -\text{arcsin}(\frac{\mu}{\sqrt{(n+\frac{1}{2})^2+\mu^2}})$. Using the approximation $\sum_{n=1}^{\infty} f(n) \approx \int_1^{\infty} f(x) dx$ for slowly changing functions (which applies to $\frac{1}{x}$ when $x \geq 1$) we get
\begin{equation}
    \text{arg}[\Gamma(\frac{1
}{2}-i\mu)] \approx \gamma \mu + \text{arcsin}\frac{\mu}{\sqrt{\frac{1}{4}+\mu^2}} + \int_1^{\infty} \Big{(}\text{arcsin}\frac{\mu}{\sqrt{(x+\frac{1}{2})^2+\mu^2}} - \frac{\mu}{x}\Big{)} dx
\end{equation}
 Expanding $\text{arcsin}(\frac{\mu}{\sqrt{(x+\frac{1}{2})^2}+\mu^2})= x' \text{arcsin}(\frac{\mu}{\sqrt{(x+\frac{1}{2})^2+\mu^2}})$ we get $$\int \text{ arcsin}(\frac{\mu} {\sqrt{(x+\frac{1}{2})^2+\mu^2
}}) \, dx = x \cdot \text{arcsin}(\frac{\mu}{\sqrt{(x+\frac{1}{2})^2+\mu^2}})+\frac{\mu}{2}\text{ln}(\mu^2+(x+\frac{1}{2})^2)+C$$
 And, therefore
 \begin{equation}
     \text{arg}[\Gamma(\frac{1
}{2}-i\mu)] \approx \gamma \mu + \mu - \frac{\mu}{2}\text{ln}(\mu^2+\frac{9}{4}) +\text{arcsin}\frac{\mu}{\sqrt{\frac{1}{4}+\mu^2}}-\text{arcsin}\frac{\mu}{\sqrt{\frac{9}{4}+\mu^2}}
\label{eq: arg(Gamma) aproximation}
 \end{equation}
Now, let us look at two asymptotics $|\mu| \gg 1$ and $|\mu| \ll 1$. Expanding \ref{eq: arg(Gamma) aproximation} for both cases we get 
\begin{equation}
    \begin{split}
        \phi(\mu) \approx \gamma \mu \, &\text{, when } |\mu| \gg 1 \\
        \phi(\mu) \approx (\gamma+\text{ln}\frac{3}{2}) \mu & +\mu \text{ln}|\mu| \, \text{, when } |\mu| \ll 1 \\
    \end{split}
\end{equation}
The equation \ref{quantonisation with negative delta} turns into
\begin{equation}
    \begin{split}
        \cos(\frac{c}{eB}S + \gamma \mu) & \approx 1 \, \text{, when } |\mu| \gg 1 \\
        \cos(\frac{c}{eB}S + (\gamma+\text{ln}\frac{3}{2}) \mu +\mu & \text{ln}|\mu|) \approx \frac{1}{\sqrt{2}}-\frac{\pi}{2}\mu \, \text{, when } |\mu| \ll 1 \\
    \end{split}
\end{equation}
\bibliography{semi_Dirac_bib}
	
\end{document}